\documentclass[pdflatex,sn-mathphys-num]{sn-jnl}

\usepackage{graphicx}%
\usepackage{multirow}%
\usepackage{amsmath,amssymb,amsfonts}%
\usepackage{amsthm}%
\usepackage{mathrsfs}%
\usepackage[title]{appendix}%
\usepackage{xcolor}%
\usepackage{textcomp}%
\usepackage{manyfoot}%
\usepackage{booktabs}%
\usepackage{algorithm}%
\usepackage{algorithmicx}%
\usepackage{algpseudocode}%
\usepackage{listings}%
\usepackage{chemformula}

\begin{document}

\title[Density super-resolution]{Image Super-resolution Inspired Electron Density Prediction}


\author[1]{\fnm{Chenghan} \sur{Li}}\email{chhli@caltech.edu}

\author[2]{\fnm{Or} \sur{Sharir}}\email{or@sharir.org}

\author[2]{\fnm{Shunyue} \sur{Yuan}}\email{syuan@caltech.edu}

\author*[1]{\fnm{Garnet K.} \sur{Chan}}\email{gkc1000@gmail.com}

\affil[1]{\orgdiv{Division of Chemistry and Chemical Engineering}, \orgname{California Institute of Technology}, \orgaddress{\city{Pasadena}, \postcode{91125}, \state{CA}, \country{United States}}}

\affil[2]{\orgdiv{Division of Engineering and Applied Sciences}, \orgname{California Institute of Technology}, \orgaddress{\city{Pasadena}, \postcode{91125}, \state{CA}, \country{United States}}}

\abstract{
Drawing inspiration from the domain of image super-resolution, we view the electron density as a 3D grayscale image and use a convolutional residual network to transform a crude and trivially generated guess of the molecular density into an accurate ground-state quantum mechanical density.  
We find that this model outperforms all prior density prediction approaches. 
Because the input is itself a real-space density, the predictions are equivariant to molecular symmetry transformations even though the model is not constructed to be.
Due to its simplicity, the model is directly applicable to unseen molecular conformations and chemical elements. We show that fine-tuning on limited new data provides high accuracy even in challenging cases of exotic elements and charge states. Our work suggests new routes to learning real-space physical quantities drawing from the established ideas of image processing.
}


\maketitle

\section{Introduction}\label{sec1}

Computing electronic properties of the ground-state of molecules and materials is a central task in chemistry and materials science.
Even within the common mean-field formulation known as Kohn-Sham density functional theory (DFT)~\cite{parr1994density}, the computations can be prohibitively expensive for many applications, for example, in molecules with many atoms, or when many geometric configurations must be considered. 
To reduce this cost,
machine learning techniques have risen to the fore. These 
 substitute physics-based calculations with a machine learning model, turning the electronic structure problem into a regression task to map a chemical description to the electronic state and properties\cite{kulik2022roadmap,ceriotti2021machine}. 

 The most common machine learning models for electronic structure target the electronic energy.
In one approach, known as machine learning force fields, the inputs are the molecular geometry and element types (and sometimes atomic energies)\cite{behler2007generalized,schutt2017schnet,bartok2010gaussian,unke2019physnet,zhang2018deep,smith2019approaching,smith2017ani,gasteiger2020fast,tholke2021equivariant,batzner20223,musaelian2023learning,batatia2022mace}. 
Another common class of models requires first carrying out a simpler molecular electronic structure calculation (e.g. semi-empirical tight-binding density functional theory) to generate an input to a model that predicts the energy of a more sophisticated electronic structure method on the same molecule~\cite{welborn2018transferability,qiao2022informing,ramakrishnan2015big}. 
In both approaches, much effort has focused on improving the featurization and model architectures, for example to 
explicitly incorporate the equivariance and invariance that arise from physical symmetries.\cite{batzner20223}.
These ML models have achieved considerable success, greatly reducing the time to obtain quantum mechanical quality energies. However, the energy is only a single facet of the electronic structure, and machine learning other properties of the quantum state has been comparatively less explored~\cite{ceriotti2022beyond}.

Within density functional theory, due to the Hohenberg-Kohn theorem~\cite{hohenberg1964inhomogeneous}, the ground-state electron density formally yields all properties of the associated quantum state. 
Machine learning the electron density has thus attracted recent attention~\cite{grisafi2018transferable,zepeda2021deep,lewis2021learning,fabrizio2019electron,mahmoud2020learning,kamal2020charge,gong2019predicting,del2023deep,chandrasekaran2019solving,fiedler2023predicting,qiao2022informing,sinitskiy2018deep}. Like their counterparts in force fields, ML density models have been explored with different sources of input data, ranging from
ones that just take in the geometry and element types~\cite{grisafi2018transferable,lewis2021learning,fabrizio2019electron,mahmoud2020learning,kamal2020charge,gong2019predicting,del2023deep,chandrasekaran2019solving,fiedler2023predicting} to ones that first require a quantum mechanical calculation~\cite{qiao2022informing,sinitskiy2018deep}. 

We will demonstrate that a powerful way to learn an accurate molecular electron density is to start from its crudest physical approximation, the superposition of atomic densities, represented on a spatial grid. In fact the atomic densities can be chosen to be extremely crude, and serve mainly to denote the size and location of the atoms, but importantly once the choice is made, they do not require recalculation for any molecule. After featurizing the input superposed atomic density on a coarse grid, our model learns the accurate molecular electron density on a high-resolution grid. This learning task is the 3D analog of image super-resolution, a well-established domain in computer vision\cite{wang2020deep,chen2022real}, where crude, lossy images at low resolution are enhanced to accurate, high-resolution images. We therefore adopt a standard image processing approach, using a simple convolutional residual neural net (ResNet)~\cite{he2016deep} that does not enforce any physical symmetries.
Despite this, it is easily argued that the predictions of our model are trivially invariant to symmetry transformations of the grid and molecule (as is targeted in an equivariant model construction~\cite{jorgensen2022equivariant}) due to the properties of our input data. 
We find that we achieve 
state-of-the-art accuracy in density prediction, improving on all previous models. 
Furthermore, because we do not use explicit element types, our architecture generalizes, and can be easily fine-tuned, across chemical space, which we demonstrate in molecules including elements we do not train against. From the density, we compute energies and orbitals through a single diagonalization of the Kohn-Sham Hamiltonian based on our predicted density. We find that even though we do not train against these quantities, the accuracy of our predictions exceeds that even of previous models that were trained against this data. We propose that the strategy in this work not only provides an important step forward in the density prediction problem, but opens up a new class of methods to predict real-space electronic structure properties based on well-established image processing techniques.

\section{Results}\label{sec2}

\subsection{Density Learning and the superposition of atomic densities}

The main objective of image super-resolution is to establish a mapping from a low-resolution and potentially inaccurate image (e.g. generated by lossy compression), denoted $\mathbf{x}$ to its high-resolution counterpart, $\mathbf{X}$. We view density learning through the same lens. Namely, given an inaccurate and low-resolution molecular density, we wish to predict the accurate molecular density, here defined as the self-consistent pseudopotential DFT density, on a finer-grid. The density learning essentially performs two tasks: it transforms the ``inaccurate'' density to an accurate one, and it upscales the spatial resolution of the density. The latter is similar to interpolation, and can be achieved without machine-learning techniques. The former is a more complex map, and is where data driven approaches are most useful.

Since we wish to learn accurate densities for large molecules, it is essential for our input data to be cheap to generate. This suggests we should avoid approaches where we must first perform a low-level molecular quantum mechanical calculation, as these will be prohibitive in large problems. On the other hand, the list of atomic types and positions alone seems more parsimonious than necessary. We will define our input data to be the physically motivated approximation of the superposition of (spherical) neutral atomic densities (SAD), normalized to the total number of electrons, expressed on a low-resolution uniform grid.
The spherical neutral atomic densities are obtained from filling electrons into atomic natural orbitals according to the Aufbau principle, which need to be tabulated once for the periodic table, while the cost of assembling the SAD guess on the the molecular grid requires no additional quantum calculation, is linear in the molecular size, and scalable to large molecules. The SAD guess preserves important physical constraints (e.g. it is positive) and provides qualitative information about the size and location of the atoms, but otherwise does not provide an accurate density, as quantified below.

\subsection{Training data and model}

Our training data will consist of DFT densities computed in the Gaussian-Plane-Wave approach~\cite{vandevondele2005quickstep,sun2018pyscf,sun2020recent} using the Perdew-Burke-Ernzerhof (PBE) functional~\cite{perdew1996generalized}, the Goedecker-Teter-Hutter (GTH) pseudopotential~\cite{krack2005pseudopotentials}, and the GTH-TZV2P basis set~\cite{vandevondele2005quickstep}, computed on a fine uniform grid, as implemented in the PySCF~\cite{sun2018pyscf,sun2020recent} package. The relation between the coarse grid for the input density and predicted fine uniform grid is a factor of 2 or 4 in each dimension as specified below (an overall factor of 8 or 64 in the number of density points; note, the high-resolution grid is fixed, so the coarse-grid is scaled).

The model is a convolutional residual neural net (ResNet), adapted from the ResNet part of the SRGAN superresolution model\cite{ledig2017photo} by extending it to 3D, and with other adaptations discussed in the SI section 1. Note that, as we do not target human perception of the visual quality of the image,
 we do not use an adversarial network for our loss function (as used in image-processing tasks to produce images that are visually realistic to humans~\cite{ledig2017photo}) but instead use a standard mean absolute error (MAE) loss
\begin{align}
    \mathrm{Err}_\rho = \frac{1}{N_e} \int \mathrm{d}\mathbf{r} |\rho^\mathrm{pred}(\mathbf{r})-\rho^\mathrm{ref}(\mathbf{r})|.
\end{align}

Recent learning approaches for molecular systems have favoured the incorporation of physical symmetries into the model, such as equivariance under the coordinate transformations corresponding to translations, rotations, and reflections. For example, in~\cite{jorgensen2022equivariant}, an equivariant network is used to ensure that the density prediction is invariant under simultaneous coordinate transformations of the nuclear positions and the grid-points at which the density is evaluated. 
Our approach achieves the same equivariance in a straightforward manner without needing to use an equivariant model architecture. This is because the input SAD density is invariant under simultaneous coordinate transformations of the molecule and grid points. When, for example, the molecule is rotated, the input density is rotated and expressed on a set of grid points which are similarly rotated. The values of the input density are thus unchanged, and since the model uses only the values of the density (not the coordinates of the grid points) the predictions are unchanged.

\subsection{Density learning on the test set}

In Table \ref{tab:density_error}, we summarize the accuracy of our density model trained on the QM9 dataset, which contains 134K small neutral organic molecules composed of the CHONF elements\cite{ramakrishnan2014quantum}. 
As is usual, the dataset is split into training (used to optimize the model parameters), validation (used to determine the end-point of training), and test (used for independent assessment) datasets, and we follow the splitting used in Ref.~\cite{jorgensen2022equivariant}.

We consider two types of input densities. The first is the standard low quality SAD input density (similar to the atomic guess approach in NWChem\cite{apra2020nwchem} but without building and diagonalizing a molecular Fock), based on pre-tabulated atomic natural orbitals expanded in a Gaussian atomic basis~\cite{widmark1990density,roos2004relativistic,roos2004main}. The other is of even lower (``atrocious'') quality, and is generated by modifying the SAD density by multiplying the Gaussian basis exponents of the atomic natural orbitals to be filled by electrons by two. This results in a highly spatially contracted density that is very poor in terms of $\mathrm{Err}_\rho$ but which still 
preserves qualitative information about the atom positions and the relative order of atom sizes.

As shown in Table~\ref{tab:density_error}, starting from the SAD guess, and using an upscaling factor of 2, our model refines the density error of the input SAD density by two orders of magnitude.
The atrocious SAD density has a worse input density error by a factor of 2 from the SAD density ($\mathrm{Err}_\rho=27.7\%$). However, using this input, the model achieves a 173-fold factor of improvement, and the predicted error in the density is then almost the same as when using the standard SAD input (model prediction $\mathrm{Err}_\rho=0.16\%$ on the test set), testifying to the robustness of the model. If we upscale the resolution by a factor of 4, the result is only slightly less accurate than upscaling by a factor of 2, with $\mathrm{Err}_\rho=0.19\%$. 

To place our results in perspective, we compare to the density prediction error of some recent density learning approaches on the same datasets. 
The invDeepDFT and eqDeepDFT models~\cite{jorgensen2022equivariant} are respectively invariant and equivariant models that take the molecular structure and element types as inputs, while OrbNet-Equi~\cite{qiao2022informing} uses data from a molecular semi-empirical tight-binding DFT calculation. We reproduce their reported errors in Table \ref{tab:density_error} and we find that we outperform all these prior models.

We further compare to the error from a direct numerical fit of the target density. For this we use Gaussian density fitting, a common numerical representation in quantum chemistry which expands the density in an auxiliary Gaussian basis~\cite{whitten1973coulombic}. (Certain density learning approaches aim to learn the coefficients of expansion of the density in such a basis~\cite{grisafi2018transferable,lewis2021learning,fabrizio2019electron,mahmoud2020learning,qiao2022informing}). Here we choose the auxiliary basis to be the default even tempered Gaussian auxiliary basis in PySCF for the GTH-TZV2P basis set, and fit using the overlap metric~\cite{mintmire1982fitting} (to minimize the least squares density error). Our density model outperforms even this direct numerical fit to the data by a factor of 2.

\begin{table}
	\caption{Density prediction error  $\mathrm{Err}_\rho(\%)$ on QM9 test set. ResNet (This work) starts from the SAD guess and uses an upscaling factor of 2. More details and other data in the text. Note that the DeepDFT, OrbNet-Equi models, and this work were fitted to QM densities generated from different DFT computational procedures, and the errors were computed with respect to each of their ground-truth densities.}
	\centering
	\begin{tabular}{llllll}
        \toprule
		ResNet (This work)    & inv-/eqDeepDFT\cite{jorgensen2022equivariant} & OrbNet-Equi\cite{qiao2022informing} & Density Fitting & SAD guess \\
		\midrule
		0.14  & 0.36/0.27 & 0.21 &    0.29 & 13.9 \\
		\bottomrule
	\end{tabular}
	\label{tab:density_error}
\end{table}

\subsection{Energies and properties from predicted densities}

Within the framework of Kohn-Sham density functional theory, the total electronic energy and orbitals can
be obtained from the exact electron density via a single diagonalization
of the effective Kohn-Sham Hamiltonian (``one-step'' DFT). The quality of the energy and orbitals therefore report on the exactness of the predicted densities by our model in a physically meaningful way, even though our model is not trained on the energy and orbitals directly. We note, however, that it can be argued that the required diagonalization in the one-step DFT means that this is not a pure data-driven approach to obtain these quantities.

\begin{table}[]
    \caption{Electronic property prediction errors on QM9. Quantities defined in the text. Energy errors ($E^\text{elec}$, $\epsilon^\text{HOMO}$, $\epsilon^\text{LUMO}$, $\Delta \epsilon$) are given as the MAE. 
    ResNet (This work) starts from the SAD guess and uses an upscaling factor of 2.}
    \centering
    \begin{tabular}{llccccc}
    \toprule
    Quantity & Unit & ResNet (This work) & Allegro\cite{musaelian2023learning} & OrbNet-Equi\cite{qiao2022informing} & PaiNet\cite{schutt2021equivariant} & ET\cite{tholke2021equivariant} \\
    \midrule
    $E^\mathrm{elec}$ & meV & 3.0 & 4.7 & 3.5 & 5.9 & 6.2 \\ 
    $\epsilon^\mathrm{HOMO}$ & meV & 12.0 &  & 9.9 & 27.6 & 20.3 \\
    $\epsilon^\mathrm{LUMO}$ & meV & 9.5 &  & 12.7 & 20.4 & 17.5 \\
    $\Delta\epsilon$ & meV & 11.0 &  & 17.3 & 45.7 & 36.1 \\
    \bottomrule
    \end{tabular}
    \label{tab:energy_error}
\end{table}

In Table \ref{tab:energy_error}, we show the MAE in energy on the QM9 dataset. We compare to predictions from a number of recent equivariant models which are trained to the energy rather than the density.
The learned densities from our model allow us to achieve state-of-art accuracy in the energy as well (out-performing all other models in the Table) even though the energy is not a training target. This attests to centrality of the density as a learning target.

The errors in the highest-occupied-molecular-orbital (HOMO) and lowest-unoccupied-molecular-orbital (LUMO) energies ($\epsilon^\text{HOMO}$, $\epsilon^\text{LUMO}$), HOMO-LUMO gap ($\Delta \epsilon$) are also reported in Table \ref{tab:energy_error}. 
As is evident, the one-step DFT HOMO/LUMO energy predictions from our predicted densities are larger than for the total energy. This is because near the DFT electronic minimum, the energy error is quadratic in the density error (which is here small), while the relationship for other observables is linear.\cite{grisafi2022electronic}. 
On the other hand, even though we did not train to these quantities, we still see competitive (or even better) accuracy for predicting HOMO-LUMO energies and the gap compared to prior models, even though these other models were trained on the HOMO-LUMO data.

\subsection{Transferability on Unseen Geometries and Elements}

We now assess the transferability of our model. As a first task, we consider the performance of our ResNet model trained on QM9 
on a different dataset of the isomers \ch{C7O2H10}. Unlike QM9 which consists only of equilibrium molecular geometries, the \ch{C7O2H10} dataset contains diverse conformers sampled at high temperature, and thus the geometries and associated electron densities extend outside of the training space of QM9. 
Specifically, the \ch{C7O2H10} dataset comprises 5,000 conformations for each of the 113 isomers of \ch{C7O2H10}. As these conformations were sampled at 1 fs intervals from molecular dynamics (MD) simulations and data at adjacent time steps are very correlated, we downsampled by a factor of 100 for each isomer trajectory, resulting in a total of 5650 less correlated data points.

As shown in Table \ref{tab:transferability}, our ResNet model trained on QM9 (denoted Zero-shot) moderately decreases in accuracy when evaluated on the \ch{C7O2H10} dataset, illustrating a qualitative difference between equilibrium and non-equilibrium structure densities. This hypothesis is supported by the high accuracy we observe when evaluating our model densities only for the initial conformation of each of the 113 isomer MD trajectories ($\mathrm{Err}_\rho=0.13\%$), which correspond to local energy minima. We also assess the Zero-shot eqDeepDFT (trained on QM9) model. This gives a zero-shot density error of 1.8\% on the full diluted dataset and 0.29\% on the first conformation of each isomer. The zero-shot error of ResNet is not only smaller than that of eqDeepDFT, but the increase from the QM9 test error is also smaller than that of eqDeepDFT. This shows that our density model not only fits better but also generalizes better.

\begin{table}[]
    \caption{Transferability check of the ResNet model trained on QM9. The errors are reported as MAE. Zero-shot means the model trained on QM9 is tested on the \ch{C7O2H10} and MT datasets, fine-tuned means that after training on QM9, the ResNet model is further fine-tuned on a small portion of each dataset.
    From-scratch represents a ResNet model trained only on the \ch{C7O2H10} or MT data, starting from randomly initialized parameters.}
    \centering
    \begin{tabular}{llcc}
        \toprule 
         Dataset & & $\rho(\%)$ & $E^\mathrm{elec}$(meV) \\
         \midrule
        \multirow{4}{*}{\ch{C7O2H10}} & Zero-shot ResNet & 0.65 & 24.5 \\
                                      & Fine-tuned ResNet & 0.25 & 2.6 \\
                                      & From-scratch ResNet & 1.5 &  \\
                                      & SAD Guess & 13.8 &  \\
                                      & Zero-shot eqDeepDFT & 1.8 &  \\
                                      & DTNN\cite{schutt2017quantum} &  & 74 \\
                                      & SchNet\cite{schutt2017schnet}\footnotemark[1] &  & 16 \\
        \cmidrule{2-4}
        \multirow{3}{*}{MT} & Zero-shot ResNet & 2.1 &  \\
                                      & Fine-tuned ResNet & 0.33 &  \\
                                      & From-scratch ResNet & 3.4 &  \\
                                      & SAD Guess & 11.9 &  \\
        \bottomrule
    \end{tabular}
    \footnotetext[1]{Trained and tested on ISO17\cite{schutt2017schnet}, containing 129 isomers, rather than the 113 isomers in the \ch{C7O2H10} data set in this work.}
    \label{tab:transferability}
\end{table}
%
%
%

Despite the decrease in accuracy of our predicted densities, when fed into a one-step diagonalization, our Zero-shot ResNet model still predicts energies within ``chemical accuracy'' of 1 kcal/mol (24.5 meV = 0.56 kcal/mol) in the total energy. Since we did not train to energies and the density learning was trained without conformational diversity of this isomer dataset, this illustrates surprising robustness of the model. In fact, the energy accuracy far exceeds that of one machine learning force field (DTNN) and is comparable to that of another (SchNet) that were specifically trained to reproduce the energy on this dataset. 

To obtain better performance,  we further fine-tune our ResNet model after training on QM9, using a small dataset comprising the first two downsampled conformations---including the one at the energy minimum---from each of the 113 trajectories. During training, we monitored the validation loss on the third conformation of each trajectory. Since the conformations in the dataset are ordered by the MD simulation time, the use of the first 3 conformations for training and validation limits our model's fine-tuning to future conformations. 

In Table \ref{tab:transferability}, we show the accuracy of the fine-tuned model on the last 40 downsampled conformations of each trajectory. As such, the middle data points (4th-9th) of each trajectory are never used, and this gap between the training/validation set and test set helps minimize the correlation between the training and test data, creating a more stringent test. We note that using the first two conformations of each isomer corresponds to only 4\% of our downsampled \ch{C7O2H10} dataset (and 0.04\% of the original dataset). Typically, such limited training data ($113\times2=226$ data points) does not suffice to train a deep neural network, thus a fine-tuning approach is essential. Training to this small dataset alone yields the From-scratch ResNet model which is less accurate than the predictions from the zero-shot ResNet model. Fine-tuning leads to a (more than) two-fold decrease of the density and energy error on the \ch{C7O2H10} set and a performance close to that achieved by ResNet in the QM9 testing set.

We next evaluate the ResNet model transferability on unseen elements and molecular charges. These are non-trivial tasks for models that explicitly require the element types as input, and (indeed, without any modification) is impossible for certain models. However, our network can be applied without any modification in both these cases. 

As a first test, we create a training and validation set for QM9 that deliberately omits the N atom, and train a model 
 to this limited data. 
The model yields a density error of $1.4\%$ when tested on the remaining nitrogen-containing QM9 molecules. This error is significantly higher than what we saw in the previous transferability tests on the isomer dataset. However, the $1.4\%$ error still represents a substantial improvement over the initial SAD guess, which has a density error of $13.6\%$. Further analysis reveals that the density error specifically associated with nitrogen is $3.3\%$, compared to a $7.8\%$ error in the SAD guess around the nitrogen atoms. (The density error around the atoms was evaluated using grid points within the atomic radii as taken from Ref.~\cite{pyykko2009molecular}).
This indicates that the pre-training on the limited QM9 dataset without nitrogen enhances the model's ability to predict densities for the unseen nitrogen element.

As a second even harder test, we consider a large biological system, namely, a model of the GTP hydrolysis reaction in microtubules\cite{beckett2023} (denoted here the MT dataset). This consists of on average 211 atoms 
and a wide range of conformations 
involving close interactions between GTP, water, and protein amino acid residues, and bond-breaking during GTP hydrolysis. 
From the density prediction perspective, the system is especially challenging as it carries a total charge of -4 and also includes
two new elements compared to the QM9 dataset, namely phosphorus and magnesium.

\begin{figure}
    \centering
    \includegraphics[width=13cm]{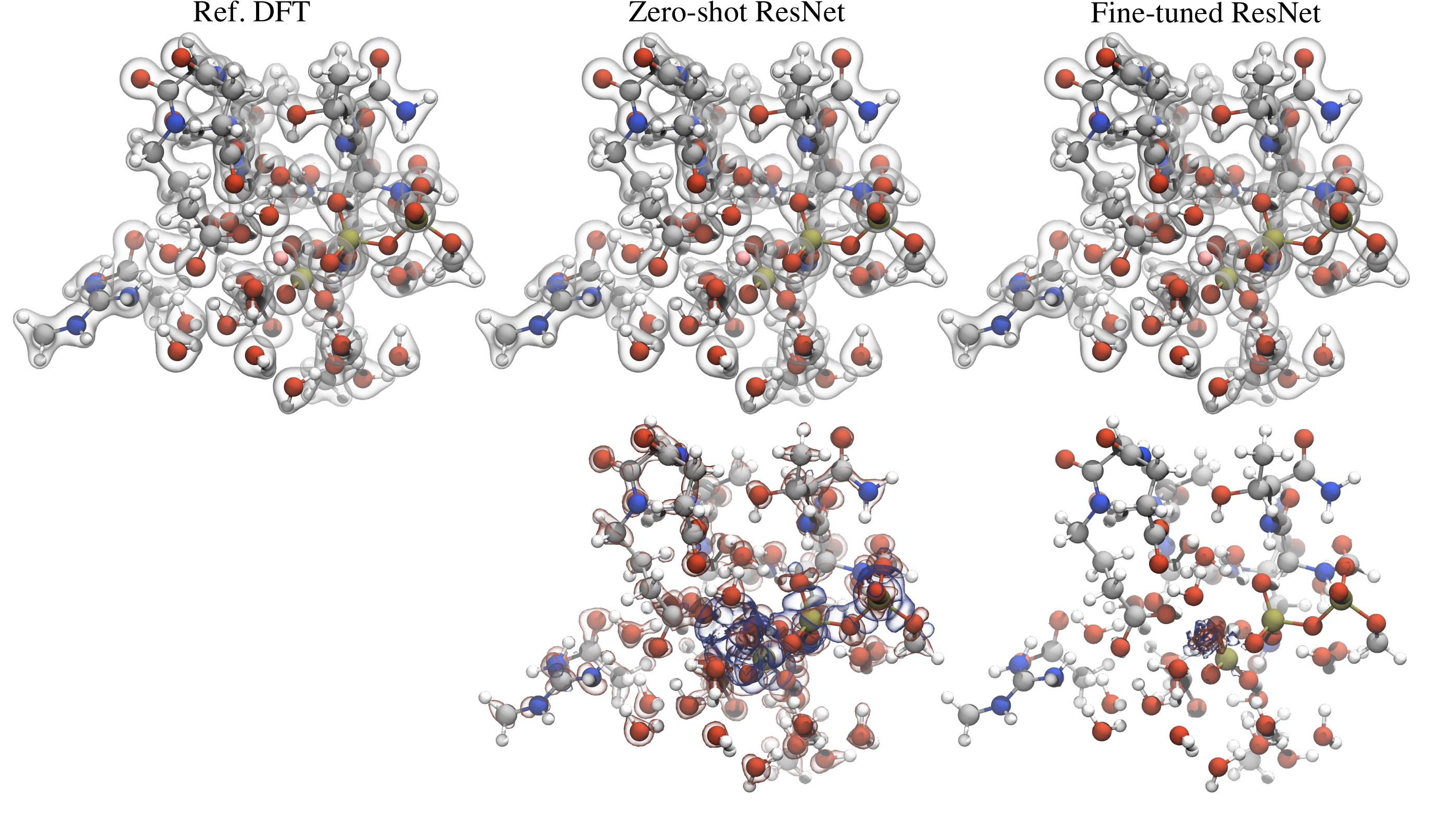}
    \caption{Density prediction on a microtubule-mediated GTP hydrolysis model. The first row shows the electron density isosurface at 0.1 $e$/Bohr$^3$. The second row shows the density prediction error ($\rho^\mathrm{pred}-\rho^\mathrm{ref}$) isosurfaces at 0.003 $e$/Bohr$^3$ (transparent red) and -0.003 $e$/Bohr$^3$ (transparent blue). Coloring of atoms: oxygen (red), nitrogen (blue), carbon (silver), hydrogen (white), phosphorus (gold), and magnesium (pink). }
    \label{fig:mt}
\end{figure}

The previous transferability task omitted nitrogen in the QM9 dataset but trained on data containing oxygen and fluorine elements. Thus testing on molecules with the unseen nitrogen element tests the interpolation ability of the model between two adjacent elements in the periodic table. In contrast, moving from the QM9 dataset to the MT dataset involves extrapolating elemental and charge state behaviour.
As such, very accurate predictions are not anticipated in this scenario. Indeed, we find using the Zero-shot ResNet model a relatively large average error ($2.1\%$) with a large density error around the phosphorus and magnesium elements  ($7.8\%$ and $33\%$ respectively). A visual check confirms that the density errors are predominantly around the unseen elements (Figure~\ref{fig:mt}). 

To improve on the Zero-shot model, we can use fine-tuning. Again, we limit ourselves to a small training set, containing a few initial conformations of the MD trajectories. Despite using limited training data (110 data points, comprising only 8\% of the total data), the fine-tuned ResNet model achieves a 0.33\% accuracy in density prediction on the test data (approximately 88\% of the entire MT dataset), again approaching the accuracy of our state-of-the-art model on the simpler QM9 training dataset. In comparison, training from scratch with the same limited data results in a model that is much less accurate than the fine-tuned one. This highlights that pre-training our density model on a standard dataset such as QM9, in conjunction with a modest amount of fine-tuning, 
achieves quantitative accuracy even in the extremely challenging case of unseen systems with new charge states and unusual elements.

\section{Discussions}
We have demonstrated that a simple machine learning strategy, 
inspired by image super-resolution methods, is a promising approach for electron density prediction. The model is based on a convolutional residual network, and the input is a crude guess of the density (superposition of atomic densities; SAD) represented on a coarse 3D real-space grid. 
We showed that even extremely poor input atomic guesses make virtually no difference to the accuracy, thus the input density mainly encodes the atomic positions and atom types (the latter implicitly represented the size of the atomic density).
The model then transforms the initial guess density into the accurate quantum mechanical ground-state density on a fine grid. The model achieves equivariance under standard molecular transformations by virtue of the properties of the input density.

From a computer vision perspective, the above process mimics the well-studied image super-resolution problem. From a physical perspective, the model may be viewed as indirectly learning the Hohenberg-Kohn (HK) map from the nuclear electrostatic potential $v^\mathrm{ext}(\mathbf{r})$ to the ground-state density, where the nuclear potential is encoded in the features of the input density. The latter may be seen as in the spirit of Bright-Wilson's argument~\cite{tozer1996exchange,rich1968structural} that the nuclear potential of the Hamiltonian may be identified from the features of the density.

Following the density prediction, a one-step DFT calculation enables the extraction of various DFT electronic properties, such as the total energies and molecular orbitals. We found that this gave state-of-art accuracy in predicting total and orbital energies 
even without training against such data. Importantly, our model also displayed significant transferability to previously unseen molecular geometries and moderate transferability for ``interpolated'' chemical elements (with atomic number in between elements seen during training). After fine-tuning our pre-trained model on limited new data corresponding to new geometries and elements, we could achieve accurate predictions in all the cases we examined.
Such transferability and fine-tunability suggests our approach is well-suited to an ensemble-based active learning setup\cite{beluch2018power,smith2018less} on large unlabeled datasets. Given its transferability, there should be a significant reduction in the need for expensive DFT data labeling. 

The success of our real-space approach to learn densities also suggests applications to other objects which are naturally expressed on real-space grids. In addition, while the electron density provides access to the electronic observables such as the energy after a single diagonalization, as performed in this work, it is desirable to use a data-driven approach to bypass this expensive computational step. These directions will be explored in the future.

\backmatter

\bmhead{Supplementary information}
Supplementary information on the computational details for data generation, and training details.

%
%

\section*{Declarations}

\begin{itemize}
\item  This work was primarily supported by the United States Department of Energy, Office of Science, Basic Energy Sciences, Chemical Sciences, Geosciences, and Biosciences Division, FWP LANLE3F2 awarded to Los Alamos National Laboratory under Triad National Security, LLC (`Triad') contract grant no. 89233218CNA000001, subaward C2448 to the California Institute of Technology.
    Additional support for GKC was provided by the Camille and Henry Dreyfus Foundation via a grant from the program ``Machine Learning in the Chemical Sciences and Engineering". 
\item The authors declare no conflict of interest.
\item The authors disclose the use of the GPT-4 (OpenAI) model during the writing of the first draft of the article. The artificial intelligence model was used to polish the language and the generated text was carefully inspected, validated, and edited by the authors. 
\end{itemize}


\bibliography{sn-bibliography}

\end{document}


\tableofcontents
\newpage

\section{Computational Details for DFT Data Generation}

All calculations employed the multigrid Gaussian plane wave (GPW) DFT as implemented in PySCF. We set a plane wave cutoff of 400 Ry, corresponding to a real-space grid spacing of approximately 0.083 \AA, and utilized the GTH pseudo-potentials and GTH-TZV2P basis sets. For all the datasets, the box size was determined based on the closest even-number multiple of the grid spacing while ensuring a minimum of 4 \AA~of space around the molecule in every dimension. This box size occasionally led to SCF convergence issues for some long conjugate chains in QM9, and a 5 \AA~margin was used instead for these cases. A SCF convergence threshold of $10^{-11}$ Hartree was set for the QM9 calculations. The GTP hydrolysis in microtubules (MT) dataset contains 1442 configurations extracted from 11 metadynamics runs\cite{beckett2023}. The total simulation time for the metadynamics trajectories was about 750 ps. Configurations spaced every 12 hours of wall time, corresponding to a 0.5-1 ps simulation interval, were included in the dataset. For the MT dataset, the SCF convergence criteria were $10^{-7}$ Hartree for the energy and $10^{-5}$ for the orbital gradient. To aid SCF convergence, we used an electronic smearing with an electronic temperature of 0.01 Hartree. All DFT calculations were conducted using the PBE functional.

\section{Model Details}

\begin{figure}
    \centering
    \includegraphics[width=13cm]{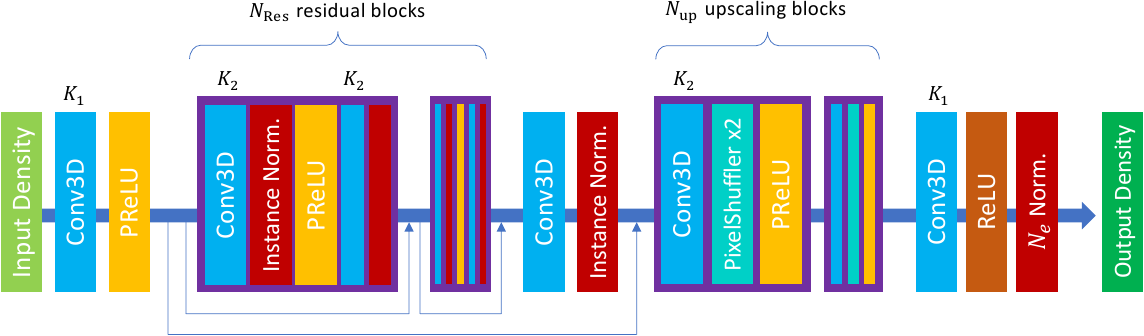}
    \caption{Model architecture. $K$ indicates the kernel size used in the convolution.}
    \label{fig:arch}
\end{figure}

The convolutional residual network employed by Ledig et al.\cite{ledig2017photo} was adapted for electron density prediction. Our modifications were (1) changing the 2D convolution into a 3D one, (2) substituting batch normalization with instance normalization, (3) appending a ReLU layer to the network output to guarantee positive density values, and (4) normalizing the ReLU output to match the correct electron number, i.e., $\rho\leftarrow N_e \rho/\int \rho \mathrm{d}\mathbf{r}$. All convolutions maintained a stride of 1, with padding applied to preserve the original size. Circular padding was employed to respect the periodic boundary conditions. The resulting architecture is depicted in Figure~\ref{fig:arch}.

\begin{figure}
    \centering
    \includegraphics[width=13cm]{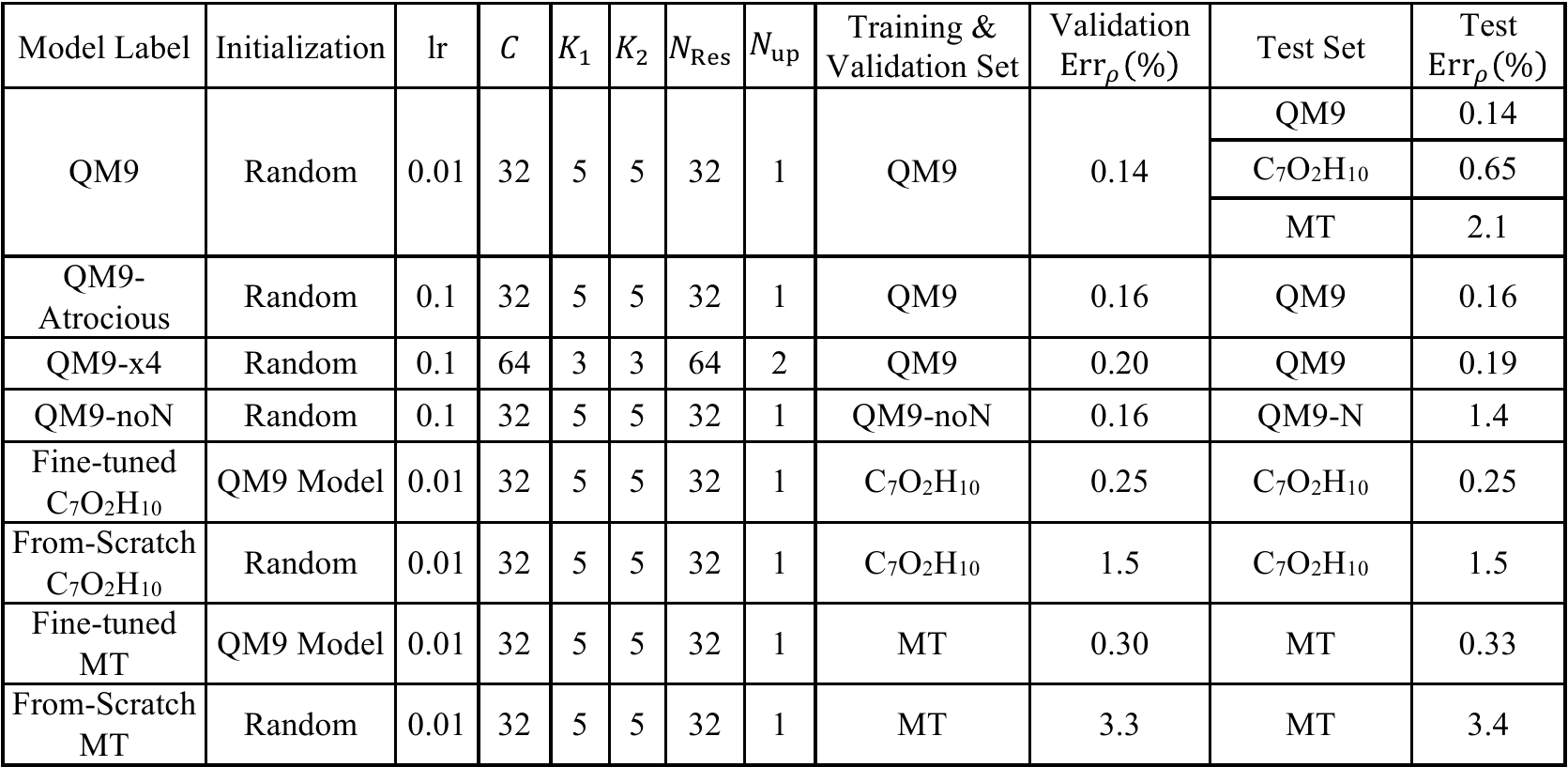}
    \caption{Density prediction error of all trained models. All the models were trained to predict the density expressed on the fine-grid (the DFT computational resolution), while the number of upscaling blocks describes the downsampling of the input density  by a factor of $2^{N_\mathrm{up}}$ in each dimension. ``lr" specifies the learning rate. $C$ indicates the channel size for the convolution. The ``atrocious" label means that the artificially worsened SAD guess was used as the model input. The ``noN" label signifies the set of QM9 molecules that do not contain a nitrogen atom, while ``N" means the remaining nitrogen-containing molecules.}
    \label{fig:density_error_all}
\end{figure}

Figure~\ref{fig:density_error_all} summarizes the hyperparameters and density errors for all models in this study. Each model underwent 50 epochs of training using the density MAE as the training loss, with the best model chosen based on the lowest validation error. The first epoch used a warm-up with a $10^5$ times reduced learning rate, followed by a cosine scheduler from the second epoch starting with a learning rate specified in Figure~\ref{fig:density_error_all}. We employed the Adam optimizer\cite{kingma2014adam} with $\beta_1=0.9$ and $\beta_2=0.99$. Training data was augmented with a random rotation along the x, y, or z axes. A batch size of 1, combined with instance normalization and data augmentation, proved sufficient for regularization across all cases, eliminating the need for weight decay. The data splitting for QM9 followed Ref.~\cite{jorgensen2022equivariant}. As mentioned in the main text, the original \ch{C7O2H10} dataset, comprising 5000 conformations for each of the 113 isomers, was down-sampled by a factor of 100 to yield less correlated data. The first two conformations of each isomer in the diluted dataset served as the training data, with the third used for validation. The entire (diluted) dataset was used for testing if the model was not trained on this dataset and otherwise the final 40 conformations were used for testing. For the MT dataset, the first 10 conformations from each of the 11 metadynamics trajectories were used for training, the 11th for validation, and the 17th to the last for testing. Fine-tuning on \ch{C7O2H10} and MT followed exactly the same training protocol, but the models were initialized from the best QM9-trained model.

\bibliography{sn-bibliography}